\documentclass[aps,showpacs,nofootinbib]{revtex4}
\usepackage{epsf,latexsym}
\usepackage{amsfonts}
\usepackage{graphicx}

\begin{document}

\title{Zero-point gravitational field equations}
\author{Alessandro Pesci\footnotetext{e-mail: pesci@bo.infn.it}}
\affiliation
{INFN Bologna, Via Irnerio 46, I-40126 Bologna, Italy}

\begin{abstract}
We study the recently reported qmetric
(or zero-point-length) expressions
of the Ricci (bi)scalar $R_{(q)}$
(namely, expressions of the Ricci scalar 
in a spacetime with a limit length $L_0$ built in),
focusing specifically on the case of
null separated events.
A feature of these expressions is that,
when considered in the coincidence limit $p \to P$, 
they generically exhibit a dependence 
on the geodesic along which 
the varying point $p$ approached $P$,
sort of memory of
how $p$ went to $P$.
This fact demands a deeper understanding
of the meaning of the quantity $R_{(q)}$,
for this latter tells about curvature of spacetime
as a whole at $P$
and would not be supposed to depend on whichever vector
we might happen to consider at $P$.
Here, we try
to search for a framework
in which these two apparently conflicting aspects
might be consistently reconciled.
We find a tentative sense
in which this could be achieved
by endowing spacetime of a specific operational meaning.
This comes, however, at the price (or with the benefit)
of having a spacetime no longer arbitrary
but, in a specific sense, constrained.
The constraint turns out to be in the form of
a relation between spacetime geometry
in the large scale (as compared to $L_0$)
and the matter content,
namely as sort of field equations.
This comes thanks to something
which happens to coincide with  
the expression of balance of (matter and spacetime) exchanged heats,
i.e. the thermodynamic variational principle
from which the field equations have been reported to be derivable.
This establishes a link between
(this specific, operational understanding of) the meaning
of the limit expression of $R_{(q)}$ on one side
and the (large-scale) field equations on the other,
this way reconnecting
(once more) the latter
to a quantum feature.
\end{abstract}


\maketitle

$ $
%
%
%
%
%
%
%
%
%
%
\section{Introduction}

When trying to combine gravity
and quantum mechanics,
a variety of results
points towards 
the existence of a lower-limit 
length $L_0$ 
(\cite{DesA, Mea, DeWA_bis, BekC-02, DeWC, 
PadA-10, PadA-09, VenA, YonA, KonA, GreA, AshA-10, RovA-02, MagA}, 
and \cite{GarA, HosA} for review and futher references).
The likelihood would then be
that the spacetime one is called to consider
ought to exhibit, in the small scale, this feature.
The approach developed in \cite{KotE, Pad01, KotI}, 
seeks precisely to implement this
through modification of the ordinary metric 
to an effective metric (also called qmetric).
It aims to provide
the specific metric framework that the spacetime 
should possess,
if it has to display 
a limit length in the small scale.

Among the encouraging results
in addressing this way
potential quantum features of spacetime 
\cite{Pad01, KotI, Pad02, Pad06, Pad12, Pad05, ChaD},
those related to the Ricci scalar $R$ can be,
due to the role this quantity plays in general relativity
as gravitational field Lagrangian,
somehow directly exploitable to shed light 
on potential quantum aspects of field equations themselves. 
In particular,
the results \cite{Pad01, KotI}
showed an expression 
for the zero-point-length Ricci (bi)scalar $R_{(q)}(p, P)$
(depending on points $P$ and $p$) 
which in the small scale ($p \to P$)
intriguingly differs, 
in the limit $L_0 \to 0$, from the value of
the (classic) Ricci scalar at the given point $P$.
This limit expression turned out to be

\begin{eqnarray}
\lim_{L_0 \to 0}  \, \lim_{p \to P} \, R_{(q)}(p, P)
=
\varepsilon D \, R_{ab} t^a t^b,
\end{eqnarray}
where
$D$ is the dimension of spacetime,
$R_{ab}$ is the ordinary Ricci tensor (at $P$),
$t^a$ is the normalized tangent vector
at $P$ to the ordinary geodesic connecting 
two space or time separated events $P$ and $p$,
and $\varepsilon \equiv g_{ab} t^a t^b = \pm 1$.     
In case of null separated events,
a recent analogous investigation has given \cite{PesP}

\begin{eqnarray}\label{isoq-2.1}
\lim_{L_0 \to 0}  \, \lim_{p \to P} \, R_{(q)}(p, P) =
(D-1) \, R_{ab} l^a l^b, 
\end{eqnarray}
with $l^a$ the ordinary null tangent vector at $P$ to the geodesic
from $P$ to $p$.  

These expressions clearly show that 
$ \lim_{L_0 \to 0}  \, \lim_{p \to P} \, R_{(q)}(p, P) \equiv R_* 
\ne R$, as mentioned.
In fact they possess an additional feature, 
the investigation of which is the focus of present study:
they exhibit an explicit dependence on the tangent 
to the geodesic at $P$,
i.e. 
$R_* = R_*(x, t^a)$ 
or
$R_* = R_*(x, l^a)$,
with $x$ concisely denoting the coordinates of $P$.
This fact raises the question of 
how we should interpret
a qmetric Ricci scalar 
at a point $P$,
if the value we could think of as assigned to it
by continuity 
along any one direction of approach, does not match with what we
find along another one;
this, strictly speaking, forbidding
to have $R_{(q)}$ smoothly defined at $P$.   
This demands for further understanding.
And as such,
this result
should be considered
not as an issue, but as a virtue.
One should 
consider it as hinting
to some deeper and as yet unveiled  
feature, of a quantum theory of spacetime. 

Other curvature-related scalars might be of help.
In particular, the Kretschmann scalar might allow
to characterize what happens
from a minimum-length standpoint
in Ricci-flat spacetimes, 
in which 
we see 
from the above 
that
the limit $R_{(q)}$ vanishes with $R_{ab}$.
This will deserve scrutiny as soon as 
a minimum-length expression for the Kretschmann
scalar will be available.

The techniques used for extracting an expression for
$R_{(q)}$ are point-splitting and coincidence-limit procedures,
similar to those used in the works aiming
to find regularized expressions for the expectation value
of stress-energy tensor $\langle T_{ab} \rangle$
on a curved background \cite{DeWA, DeWB, Xen}.
The works in this latter context,
specifically in \cite{ALN},
quantities have been considered
(intervening in the expression of $\langle T_{ab} \rangle$)
which do exhibit in principle a residual dependence 
on the separation direction
after the coincidence limit is taken,
a situation with analogies to what described here.
There, 
the need was
to have 
$\langle T_{ab} \rangle$
a definite quantity assigned at a point,
and the way to face the dependence
on the separation direction was basically
to average over the possible directions
($\langle T_{ab} \rangle$ is an
expectation value after all).  

We might try to do the same,
but the focus of present study,
meant as a first step towards a more comprehensive analysis,
is not on a ``need''
of a single-valued quantity,
but to pause and take note of the fact
that in a spacetime endowed with limit length
we do not get a single-valued quantity,
and consider it like a possible glimpse
of some possible underlying truth
one might want to try to extract.
We do this way,
prompted by the fact that
it is not at all obvious a priori
that a spacetime with a limit length
has to show such a multivaluedness for $R_{(q)}$,
and as a matter of fact
it came as a surprise
when it first was found
\cite{Pad01, KotI}.

In this vein,
in the present work
we try to gain some insight into this fact
looking at it
from the following perspective.
We ask:
Is there some sense in which
the coincidence limit of $R_{(q)}$ along some given path,
can be considered independent of $l^a$ at $P$?
Could we distinguish between path independence
of the coincidence limit of $R_{(q)}$,
a thing which clearly we mathematically do not have,
and the operational notion of
independence from $l^a$ of the value we obtain for the limit
of $R_{(q)}$ once a probe of $R_{(q)}$ 
at $P$ has been taken already? 
Is there any meaning in this?
Our hope is that actually there is,
and that something interesting could be extracted
through knowledge of
whether it is possible
to have an independence from $l^a$ of the probed $R_{(q)}$ at $P$;
and, in case of affirmative answer,
one would like to find out what this might mean.
This is what we try to do here, 
considering specifically the case of null separated events.

%
%
%
%
%
%
%
%
%
%
\section{Definition of the problem}

Let us consider a point $P$ in $D$-dimensional spacetime $M$
($D\ge 4$; metric $g_{ab}$ with signature $(-, +, + , ...)$) 
and consider the zero-point-length metric $q_{ab}(p, P)$ with base $P$
as defined for points $p$ null separated from $P$.
It reads \cite{PesN}

\begin{eqnarray}
q_{ab} = A \, g_{ab} +
\Big(A - \frac{1}{\alpha}\Big) (l_a n_b + n_a l_b),
\end{eqnarray}
where
$l^a$ is the ordinary (null) tangent, at $p$, 
to the geodesic connecting $P$ and $p$,
$n^a$ is an auxiliary null vector with 
$g_{ab} n^a l^b = -1$ and $g_{ab} n^a e^b = 0$
for any spacelike vector $e^a$ transverse to $l^a$
(i.e. $g_{ab} l^a e^b = 0$),
and
$\alpha = \alpha(p, P)$ and $A = A(p, P)$
are biscalars, functions of 
the difference of affine parameter
$\lambda(p, P) \equiv \lambda$ 
given by

\begin{eqnarray}\label{alpha}
\alpha = \frac{d\lambda}{d\tilde\lambda}
\end{eqnarray} 
and

\begin{eqnarray}\label{A}
A = \frac{\tilde\lambda^2}{{\lambda}^2} \,
  \bigg(\frac{\Delta}{\tilde\Delta}\bigg)^{\frac{2}{D-2}},
\end{eqnarray}
where
$\tilde\lambda = \tilde\lambda(\lambda)$ 
is the difference in the qmetric-affine parameter,
which has $\tilde\lambda \to L_0$ for $p \to P$.
The derivative in (\ref{alpha}) is thought as taken
at $p$,
and

\begin{eqnarray}\label{vanVleck}
\Delta(p, P) = - \frac{1}{\sqrt{g(p) g(P)}} \, 
{\rm det}\Big[-\nabla^{(p)}_a \nabla^{(P)}_b \frac{1}{2} \sigma^2(p, P)\Big]
\end{eqnarray}
($\sigma^2$ is the squared geodesic distance)
is the van Vleck determinant 
(\cite{vVl, Mor, DeWA, DeWB}; see \cite{Xen, VisA, PPV})
and

\begin{eqnarray}\label{tilde_vanVleck}
\tilde\Delta(p, P) = \Delta(\tilde p, P), 
\end{eqnarray}
with
$\tilde p \in \gamma$ such that $\lambda(\tilde p, P) = \tilde\lambda$.
In the limit $\lambda/L_0 \to \infty$,
$\alpha$ and $A$ satisfy 
$\alpha \to 1$ and $A \to 1$ \cite{PesN},
and the ordinary metric $g_{ab}$
is recovered.

In this description through the qmetric $q_{ab}$,
all the effects of the degrees of freedom of the
as-yet-unknown microscopic theory are supposed
to have been encoded in the function 
$\tilde\lambda = \tilde\lambda(\lambda)$
(this for null separations; 
a function $S = S(\sigma^2)$,
with $S$ the squared geodesic distance modified 
according to the qmetric,
analogously encodes these effects for time or space separations
\cite{KotE, KotI}). 
This function is conceived as `universal',
where we mean with this, 
any time we have a $\lambda$
we get a corresponding $\tilde\lambda(\lambda)$ irrespective
of the specific geometric characteristics
of the spacetime at the point under consideration 
or the dynamical evolution it is experiencing.
The approximation is then such that
the biscalar $\tilde\lambda$ has no proper dynamics distinct from
the (possible) dynamics of $\sigma^2$ or $g_{ab}$; 
its evolution is completely
determined by that of $g_{ab}$.
Moreover, 
the main interest within our approach
is in taking the coincidence limit $p \to P$,
and the only aspect which matters
is the fact that $\tilde\lambda \to L_0$,
with no regard to the details of how $\tilde\lambda$ reaches $L_0$.
The general idea would then be that $\tilde\lambda (\lambda)$ 
is actually determined
by degrees of freedom pertaining to the unknown description
of quantum gravity, 
but the level of our approximation here is such that
the effects of these dofs are, in a sense, meant as frozen in the 
$\tilde\lambda(\lambda)$ (we do not track their own evolution
in connection with $g_{ab}$ evolution)
when $\lambda \gg L_0$, 
and essentially amount in $\tilde\lambda \to L_0 \ne 0$
when $\lambda \to 0$.

We borrow now the expression 
for the qmetric Ricci scalar $R_{(q)}$ 
for null separated events
from \cite{PesP}:

\begin{eqnarray}\label{Rq}
R_{(q)}(p, P) &=& 
\frac{1}{A} \, R_\Sigma
- 2 \, \alpha \, \frac{d\alpha}{d\lambda} \, K
+ 2 \, \alpha^2 \, R_{ab} \, l^a l^b
- (D-2) \, \alpha \, \frac{d\alpha}{d\lambda} \, 
\frac{d}{d\lambda} \ln A
- (D-2) \, \alpha^2 \, \frac{d^2}{d\lambda^2} \ln A
\nonumber \\
&-& \frac{1}{4} (D-2) (D-1) \, \alpha^2 \, 
\Big(\frac{d}{d\lambda} \ln A\Big)^2
- \alpha^2 K^2
+ \alpha^2 K^{ab} K_{ab}
- (D-1) \, \alpha^2 \,
\Big(\frac{d}{d\lambda} \ln A\Big) K.
\end{eqnarray}
Here, 
the circumstances are assumed to be that
the Ricci scalar at $P$ is completely described
in terms of a congruence of affinely parameterized (parameter $\lambda$
for $g_{ab}$ and $\tilde\lambda$ for $q_{ab}$, with $g_{ab}$-tangent
$l^a$)
null geodesics emerging from $P$,
and the expression applies in the limit of $\lambda$ small.
$\Sigma = \Sigma(P, \lambda)$ 
is the $(D-2)-$surface 
locus of the points $p'$, each on a null geodesic from $P$ and in the 
future of it,
at the $\lambda$ corresponding to $p$, 
i.e.
$
\Sigma(P, \lambda) =
\{p' \in L: \lambda(p', P) = \lambda (>0)\}
$,
with $\lambda = \lambda(p, P)$ fixed,
where
$
L = \{p' \in M: \sigma^2(p', P) = 0, 
\ {\rm and} \ p' \ {\rm in \ the \ future \ of} \ P\}.
$
All vectors and tensors in expression (\ref{Rq}) are ordinary
--i.e. not qmetric-- vectors and tensors 
and are evaluated, as well as the scalars $R_\Sigma$ and $K$, 
 at $p$, 
and indices are lowered and
raised using $g_{ab}$ and $g^{ab}$.
$R_{\Sigma}$ is the Ricci scalar intrinsic to $\Sigma$,
$K_{ab}$ the transverse field
$
K_{ab} = {h^c}_a {h^d}_b \nabla_c l_d
$
with
$
h_{ab} = g_{ab} + l_a n_b + n_a l_b
$
the transverse metric,
and
$K = {K^a}_a$. 

When
$\lambda$ is small but
$\lambda \gg L_0$, 
we have $\alpha \simeq {\rm const} = 1$ and
$A \simeq {\rm const} = 1$,
and several terms on the rhs of (\ref{Rq}) are vanishing.
Writing, in these circumstances,

\begin{eqnarray}
\alpha = 1 + \epsilon \Phi(\lambda),
\end{eqnarray}

\begin{eqnarray}
A = 1 + \epsilon \Psi(\lambda)
\end{eqnarray}
with $\Phi$, $\Psi$ smooth functions
and $\epsilon \ll 1$ constant,
and assuming that not only the functions
$(\alpha -1)$ and $(A -1)$ are small but that also their derivatives 
of every order are small with them when 
$\lambda/L_0 \gg 1$,
what we are left with is

\begin{eqnarray}
R_{(q)}(p, P)
&=& R_\Sigma(p) + 2 \, R_{ab}(p) \, l^a(p) \, l^b(p) 
- K^2(p) + K^{ab}(p) \, K_{ab}(p) + {\cal O}(\epsilon {\cal R}),
\end{eqnarray}
with ${\cal R}$ a typical component
of Riemann tensor.
Thus, at leading order, 

\begin{eqnarray}\label{isoq-2.2}
R_{(q)}(p, P)
&=& R_\Sigma(p) + 2 \, R_{ab}(p) \, l^a(p) \, l^b(p) 
- K^2(p) + K^{ab}(p) \, K_{ab}(p)
\nonumber \\
&=&
R(p),
\end{eqnarray}
with last equality from \cite{PesP}
(equation (45) there, for $\lambda$ small)
(as for Gauss-Codazzi equations, generalized to
the case of null hypersurfaces,
see e.g. \cite{ChaC_bis} and \cite{Gem}). 

This result shows that at large scale 
(meaning this that, even if $\lambda = \lambda(p, P)$ is small, 
we have  $\lambda \gg L_0$)
the qmetric with base at $P$ gives for the Ricci scalar at $p$
an expression with no dependence on the tangent $l^a$ at $p$,
and this irrespective to the (small) value of $L_0$.
One could argue that this refers strictly speaking to $p$, not $P$,
and that $p$ can be after all also far away from $P$.
But, what we just said can anyway be used to tell
what the qmetric curvature is at $P$, precisely.
To this end, let us consider the following.
Fixed a scale, i.e. assigned a value for $\lambda/L_0$,
the qmetric geometry at $P$ can be computed using
null geodesics connecting different events $P', P'', ..$ with $P$,
each chosen to have
$\lambda'(P, P') = \lambda''(P, P'') = \lambda$ and using
as base points $P'$, $P'', ... \, .$
For the qmetric Ricci scalar at $P$,
at leading order this gives
 
\begin{eqnarray}\label{isoq-2.5}
R_{(q)}(P, P') =
R_{(q)}(P, P'') = ... =
R(P)
\end{eqnarray}
i.e.,
provided the event $P$ is reached by a null geodesic
from an event $P'$ with $\lambda(P, P') \gg L_0$,
the value of the qmetric Ricci scalar in $P$ has at leading order 
no dependence
on the chosen geodesic 
and does coincide
with the value there of ordinary Ricci scalar $R(P)$.
As such, it exhibits no dependence on the tangent $l^a$ 
to the geodesic in $P$.
We can summarize the results (\ref{isoq-2.2}) and (\ref{isoq-2.5})
by saying that
$R(P)$ gives the expression of the (minimum-length) Ricci scalar at $P$
in the {\it large scale}
with no dependence in it on the geodesic we may have used
to reach $P$.
We can write this as 

\begin{eqnarray}\label{isoq-3.4}
R_{(q)}^{(\rm Macro)}(P) = R(P),
\end{eqnarray}
having defined

\begin{eqnarray}\label{isoq-3.5}
\lim_{\lambda' \to \infty} R_{(q)}(P, P') =
\lim_{\lambda'' \to \infty} R_{(q)}(P, P'') = ... 
\equiv
R_{(q)}^{(\rm Macro)}(P).
\end{eqnarray}

In the {\it small scale},
the situation 
appears 
very different. 
What one finds from equation (\ref{Rq}) in the limit $p \to P$,
is (see appendix \ref{AppA})

\begin{eqnarray}\label{isoq-20.1}
\lim_{p \to P} R_{(q)}(p, P) 
&=&
(D-1) \, (R_{ab} l^a l^b)(P) + 
{\cal O}\bigg(\frac{L_0}{L_R} \ R_{ab} l^a l^b(P)\bigg),
\nonumber \\
&=&
(D-1) \, (R_{ab} l^a l^b)({\bar p}) + 
{\cal O}\bigg(\frac{L_0}{L_R} \ R_{ab} l^a l^b({\bar p})\bigg),
\end{eqnarray}
with $\bar p$ the event on the null geodesic
at $\lambda({\bar p}, P) = L_0$,
and $L_R \equiv 1/\sqrt{{R_{ab} l^a l^b}(P)}$ a length scale associated,
for the given $l^a$, 
with the assigned curvature at $P$.
In writing this, we assume that our ordinary spacetime 
obeys the null convergence condition, 
this then ensuring $R_{ab} l^a l^b$ is non-negative.
With $L_0$ orders of Planck length,
for ordinary curvatures
we generically assume that we are
at conditions in which
$L_0/L_R \ll 1$
(this implying to say that the event $\bar p$
at $\lambda = L_0$
is near enough to $P$ to give 
$|g_{ab}(\bar p)| = {\cal O}(R_{abcd}) \, {L_0}^2 \ll 1$ 
in a local frame (with Riemann normal coordinates)
in which $\lambda$ is length),
and that the effects of the non leading terms in equation (\ref{isoq-20.1})
result negligible
(mathematically, what we are assuming is that
$L_0$ belongs to a right neighbourhood $[0, \bar L)$ of 0
with $\bar L$ small enough that this is satisfied).
Whether this 
--at some event
in an actual spacetime--
can be appropriate or not, 
must be checked carefully,
and how to tag potentially 
non-negligible terms
is discussed in appendix \ref{AppA}.

As above,
the limiting behaviour of $R_{(q)}(p, P)$ 
can be seen as telling us what the qmetric curvature
is at $P$, this time however at a small scale.
We have just to look at null geodesics $\gamma'$, $\gamma''$, .. , 
with $g_{ab}$-affine parameters
$\lambda'$, $\lambda''$, .., 
arriving at $P$ and having started at
points $P'$, $P''$, .. with 
$\lambda'(P, P') = \lambda''(P, P'') = ... = L_0$.
This gives  

\begin{eqnarray}\label{isoq-4.2}
R_{(q)}(P, P') \ne
R_{(q)}(P, P'') \ne ...
\end{eqnarray}
in general,
with

\begin{eqnarray}
R_{(q)}(P, P') &=& (D-1) (R_{ab} {l'}^a {l'}^b)(P),
\nonumber \\
R_{(q)}(P, P'') &=& (D-1) (R_{ab} {l''}^a {l''}^b)(P), 
\\
&....&
\nonumber
\end{eqnarray}
(${l'}^a$, ${l''}^a$, .. are tangents at $P$ to the geodesics
$\gamma'$, $\gamma''$ , ..)
at leading order.
The net
result coincides with what one gets
considering equation (\ref{isoq-20.1})
in the limit $L_0 \to 0$, i.e. equation (\ref{isoq-2.1}).

What we have thus is a situation in which when the qmetric
Ricci (bi)scalar is probed (through null separations) at a large scale
($\lambda \gg L_0$) at a generic point $P$, 
its value does coincide with ordinary Ricci scalar at $P$
and, of course, does not depend on the path through which we reached $P$;
when instead we probe it at a smaller scale, potentially to the
smallest conceivable scale ($\lambda \to 0$), the qmetric Ricci (bi)scalar
deviates from its ordinary value and, moreover,
acquires a dependence 
on the geodesic path followed to reach $P$.
This memory of the path then is not present macroscopically
but appears to unavoidably arise in the small scale.

We note that
this is not something about inhomogeneity,
i.e. what might be expected 
if one could imagine spacetime as somehow granular at the small scale,
we have indeed a dependence on the direction.
It is not about anisotropy either,
for the dependence of the direction we have
is not in the sense of something we get when leaving $P$
along one direction rather than another,
but something defined {\it at} $P$,
i.e. pertaining to event $P$.
This entails that $R_{(q)}$, meant as a function, 
cannot have a small-scale smooth definition at $P$, 
for results,
it cannot be continuously prolonged at $P$;
this even if we had decided to give up, at the smallest scale, 
with the notion of `point' 
(cf. \cite{BerA}), for anyway the small-scale value
we should assign by continuity
to the Ricci scalar would depend on the direction
through which we have reached the `spot' that potentially replaces $P$.

This brings to the following consideration.
$R_{(q)}$ as defined at a point $P$
might be sort of multiple-valued entity,
expressing, from an operational point of view,
an unprobed configuration.
Different values
would correspond to different results of probes at $P$;
the difference in the values would then reflect
a difference on the results of measurements, 
not a
dependence of $R_{(q)}$ itself at $P$ on direction.
Indeed such kind of dependence
would seem hardly acceptable,
for $R_{(q)}$ at $P$ is an intrinsic geometric property
of spacetime,
and as such we can expect
it not to be dependent on whichever vector we can consider at $P$.

We see, this perspective suggests a description
which might be quantum.
More explicitly,
taking the coincidence limit along a macroscopically 
assigned geodesic
can be thought of as performing
a measurement
on the quantum system consisting of spacetime at $P$
of some quantum observable $\widehat R$
expressing the Ricci scalar.
In connection with the given macroscopic 
value of $R_{ab}$ at $P$,
the measurement on the unprobed system
is assumed to provide the result
$(D-1) R_{ab} l^a l^b$
with $l^a$ the null vector
at $P$ in the direction along which we reach $P$.
In doing so,
what is supposed to happen is that,
even if we started on a macroscopically assigned geodesic,
we actually reach $P$ microscopically 
along a specific direction chosen virtually 
at random among all directions at $P$,  
due to the uncertainty
in the momentum as we get closer and closer to $P$.
After the measurement, 
quantum mechanics requires that
the quantum system 
(spacetime at $P$) is
in an eigenstate associated 
to the eigenvalue $(D-1) R_{ab} l^a l^b$ of $\widehat R$,
still with a same macroscopic $R_{ab}$.
In any further measurement 
of the already probed Ricci scalar
slightly afterwards,
whichever is the tangent $l'^a$ 
with which we now reach $P$,
and which, were the system unprobed,
would give $(D-1) R_{ab} l'^a l'^b$,
we are required to get that same value $(D-1) R_{ab} l^a l^b$.

In this perspective, there should be some mechanism 
of quantum mechanical origin in action,
which, 
once the system has been already probed,
prevents to find as a result of 
a further measurement
something different from what already found.
In other words,
the probe at $P$ of this intrinsic geometric quantity
may well depend on $l^a$,
but once we get a value, this
should be considered as not dependent anymore
on the vector $l^a$ at $P$,
thus representing a geometric property
of spacetime as a whole there.
This might result somehow puzzling.
The following
is an attempt to make sense of it.
We go to see a potential
way this could make sense,
this going hand in hand however with
the (large scale) spacetimes we are dealing with
cannot be given arbitrarily.
This will raise the question 
of which turns out to be the relationship
between these spacetimes we get
and actual, experimentally probed, spacetime.

%
%
%
%
%
%
%
%
%
%
\section{Irrelevance of $l^a$ after a probe (empty space)}

We have independence from $l^a$ at $P$ after a probe, if
the above-defined quantity,
$R_* \equiv (D-1) R_{ab} l^a l^b$,
which has a manifest dependence on $l^a$,
i.e. $R_* = R_*(x, l^a)$,
does result,
as a consequence of constraints on $R_{ab}$, 
to be actually independent of $l^a$.
We describe this, writing

\begin{eqnarray}\label{isoq-5.1}
\frac{\partial}{\partial l^a}
\bigg[(D-1) \, R_{cd} l^c l^d + \mu \, g_{cd} l^c l^d\bigg] = 0,
\,\,\,\, {\rm at} \,\, {\rm any} \,\, l^a \,\, {\rm null.}
\end{eqnarray} 
Here $\mu = \mu(x)$ is a scalar which acts 
as a Lagrange multiplier.
Its introduction
corresponds to require that the variation is done
while keeping $l^a$ null, i.e. $g_{ab} l^a l^b =0$. 
This is to be consistent with the fact that the expression
$(D-1) \, R_{cd} l^c l^d$ is specific to the case of null separations.
We would like to emphasize that in writing (\ref{isoq-5.1})
we are not taking any directional derivative:
what we are considering is the quantity
$\Omega \equiv \lim_{p \to P} R_{(q)}$ at $P$,
with this quantity explicitly depending 
(and this is precisely the item we are addressing)
on the way $p$ approached $P$.
The varied value of $\Omega$ is always still at $P$,
and is obtained simply varying $l^a$ in the expression of $\Omega$,
forgetting how $l^a$ came about.

Equation (\ref{isoq-5.1}) gives

\begin{eqnarray}
\big[(D-1) \, R_{ac} + \mu g_{ac}\big] \, l^c = 0, 
\ \ \ \forall \, l^a \, {\rm null}.
\nonumber
\end{eqnarray}
If this is satisfied, it also is

\begin{eqnarray}
\big[(D-1) \, R_{ab} + \mu g_{ab}\big] \, l^a l^b = 0, 
\ \ \ \forall \, l^a \, {\rm null},
\nonumber 
\end{eqnarray}
i.e.

\begin{eqnarray} \label{isoq-5.5}
R_{ab} \, l^a l^b = 0,
\ \ \ \forall \, l^a \, {\rm null}.
\end{eqnarray}
This means that, looking at potential irrelevance of $l^a$
through equation (\ref{isoq-5.1}),
leaves as unique configuration
that which satisfies equation (\ref{isoq-5.5}).
Notice this gives $L_R = \infty$;
we are thus surely at conditions in which
in equation (\ref{isoq-20.1}) we have $L_0 \ll L_R$.

We can readily inspect the characteristics of this ordinary spacetime.
Equation (\ref{isoq-5.5}) implies

\begin{eqnarray}\label{isoq-6.5}
R_{ab} = \xi \, g_{ab}
\end{eqnarray}
with $\xi = \xi(x)$ a scalar.
This gives

\begin{eqnarray}\label{isoq-7.3}
G_{ab} = \Big(\xi - \frac{1}{2} \, R\Big) \, g_{ab},
\end{eqnarray}
which, from Bianchi identity, implies
$
\partial_a (\xi - \frac{1}{2} \, R) = 0,
$
namely 
$
\xi - \frac{1}{2} \, R = {\rm const},
$
and thus (\ref{isoq-7.3}) reads

\begin{eqnarray}\label{isoq-7.8}
G_{ab} = C \, g_{ab},
\end{eqnarray}
with $C$ a constant, independent of $x$
(as for the mathematical procedure 
we have followed here, cf. \cite{PadN}
exercise 15.3).

Summing up,
in order for equation (\ref{isoq-5.1}) to hold,
equation (\ref{isoq-7.8}) must hold.
We see then that
the obtaining
of irrelevance of $l^a$ after a probe,
as implemented through equation (\ref{isoq-5.1}),
demands that the ordinary, or classical, spacetime
be an Einstein space.

We note that equation (\ref{isoq-7.8}) has the nature of 
(vacuum) field equations.
Then, independence of $l^a$ after a given probe of $R_{(q)}$ at $P$
results connected to
the classical metric not being generic, but obeying instead something
which has the status of field equations.
From (\ref{isoq-7.8}),
all Einstein spacetimes, 
i.e. in particular
all vacuum solutions to Einstein equations, 
do admit a qmetric description
in which the quantum Ricci scalar at $P$
can be consistently considered (in the operational sense above)
an intrinsic geometric property of spacetime as a whole. 

We may wonder whether this
exhausts all spacetimes
which do admit such a consistent qmetric
description.
Clearly, one would not expect
this to be the case.
Retracing what we have done,
it is clear that we did not refer to any potential agent
on geometry apart from spacetime itself.
No contributors have been allowed
to determine  
the geometry of spacetime;
this is to say,
what we have considered up to now
has been spacetime devoid of any physical agency on it. 
We need, then, to look at irrelevance of $l^a$ after a given probe
also in a somehow more general context,
with matter -which is an obvious missing ingredient- 
allowed to enter the scene.

%
%
%
%
%
%
%
%
%
%
\section{Irrelevance of $l^a$ when in presence of matter}

A more general context is achieved if
we assume that, starting from the small-scale expression
$(D-1) R_{ab} l^a l^b$ for $R_{(q)}(P, P')$, 
corresponding to $R_{(q)}$ probed at $P$
through
an assigned (null) geodesic $\gamma$ with tangent $l^a$ at $P$,
any variation we can have
when we further probe $R_{(q)}$ with a changing $l^a$
(leaving it null and forgetting how $l^a$ went about)
is cancelled, or absorbed, 
by the effects of matter. 
%
This corresponds to
introduce matter as something
which acts
as needed
to endow
the small-scale Ricci scalar at $P$,
when considered operationally, 
a meaning which fits with being a quantity
determined by the geometry of spacetime as a whole at $P$,
as specified above.
Doing this, implies in particular to consider
matter as something somehow capable to affect (large-scale) geometry,
i.e. precisely what we learn from general relativity.
We express this,
associating to matter
some entity with geometric significance.
It is natural to conceive this as
a scalar geometric quantity $Q$,
much the same way as to geometry itself
can be associated the Ricci scalar $R$.
In other words, we are thinking of $Q$ as something
which parallels, as for the geometric effects of matter,
what $R$ expresses for geometry itself. 

Exactly as it happens for $R$, 
we can imagine that this geometrical quantity $Q$ has
a zero-point-length biscalar counterpart $Q_{(q)}(p, P)$ 
at generic base point $P$, 
and that, along the geodesic $\gamma$ connecting $p$ and $P$, 
$Q_{(q)}$ has the small-scale (and $L_0 \to 0$) limit

\begin{eqnarray}
\lim_{L_0 \to 0} \, \lim_{p \to P} \, Q_{(q)}(p, P) =
Q_*,
\end{eqnarray}
with $Q_*$ depending on the geodesic which goes through $P$, 
$Q_* = Q_*(\gamma)$.

Now, irrelevance of $l^a$ after a probe of $R_{(q)}$ at $P$
is introduced as follows.
We require that 
every further variation of the term $R_{ab} l^a l^a$ for the probed system,
when we slightly 
change $l^a$,
is compensated 
by an equal and opposite variation induced by matter.
In the same way,
if in presence of matter
we require $l^a$-independence of $R_{ab} l^a l^b$ alone, without taking
in due account $Q_*$, the $R_{(q)}$ we probed 
will exhibit at the end a (quite unacceptable) dependence on $l^a$.
This means that
the independence from $l^a$
after a probe of $R_{(q)}$ with matter present, 
is connected with the quantity

\begin{eqnarray}\label{isoq-8.2}
F \equiv
(D-1) \, R_{cd} l^c l^d - Q_*
\end{eqnarray}
having vanishing variation
with respect to $l^a$ for variations
which keep $l^a$ null.

When this variation is required to vanish,
it is clear that
the term
$R_{ab} l^a l^b$,
i.e. $\lim_{p \to P} R_{(q)}$,
will keep having the same dependence
on $l^a$ as before.
That is absolutely true.
Our point however is different.
As we said, what we maintain is that,
given the value $V$ that
the small-scale Ricci scalar has
at $P$
as first probed through some specific
null geodesic through which $p$ approached $P$
with tangent $l^a$ at $P$,
what must happen is,
when we change the vector $l^a$ at $P$
in the probed spacetime,
$V$
must not change anymore,
i.e. its variation must vanish.
This happens to be ensured endowing matter
with the capability to influence geometry.
Had we probed $R_{(q)}$ at $P$ for the still unprobed spacetime
with a
tangent $l'^a$ at $P$,
we would have found a different value $V'$ of the small-scale Ricci scalar;
but that value would have had
in turn to remain the same in
further probes at $P$ on the already-probed spacetime. 

Assigned $Q$, we can think of 
$Q_*$ as exhibiting:
$a)$ no dependence on $l^a$, i.e. 
$Q_* = Q_*(x)$;
$b)$ a linear dependence on $l^a$, i.e.
 $Q_* = Q_*(x, l^a) = Q_a l^a$, 
 with $Q_a$ not depending on $l^a$;
$c)$ a quadratic dependence on $l^a$, i.e.
 $Q_* = Q_*(x, l^a) = Q_{ab} l^a l^b$, 
 with $Q_{ab}$ not depending on $l^a$
 and symmetric without loss of generality;
$d)$ a cubic dependence on $l^a$, i.e.
 $Q_* = Q_*(x, l^a) = Q_{abc} l^a l^b l^c$, 
 with $Q_{abc}$ not depending on $l^a$
 and symmetric in all its indices without loss of generality;
$e)$ a higher power dependence on $l^a$;
$f)$ any combination of the above.

Before we proceed, we make a comment
on definition (\ref{isoq-8.2}).
Since 
$R_{cd} l^c l^d$ (specifically $(1/L_{Pl}^2) R_{cd} l^c l^d$)
has the physical meaning of heat density \cite{Pad20},
a same physical meaning should have $Q_*$.
We see then that
the request of irrelevance of $l^a$ at $P$ of probed spacetime
corresponds to 
the law
of balance, or equilibrium, of two heat densities.
In other words, 
that sort of logical consistency condition
we referred to as irrelevance of $l^a$ after a probe, 
is automatically satisfied
when the physical law of balance of heat densities holds true.
This accords with that,
if matter sets the geometry,
this happens
in thermodynamic terms, 
in particular as an expression
of thermodynamic equilibrium.  

Let us proceed now to discuss
the various cases above.
We immediately recognize in $(a)$ the case we have considered
in the previous section. 
This implies that case $(a)$ is equivalent to empty space.
Indeed, what we get is the same
we get with
$Q_* = Q_{(q)} = Q = 0$. 
Clearly, all the cases include in particular case $(a)$;
this happens when $Q_{a ...} = 0$.  
As for $(b)$,
imposing the vanishing of the variation of $F$
gives

\begin{eqnarray}\label{isoq-9.1}
\frac{\partial}{\partial l^a}
\bigg[(D-1) \, R_{cd} l^c l^d - Q_c l^c
+ \mu \, g_{cd} l^c l^d\bigg] = 0,
\,\,\,\, {\rm at} \,\, {\rm any} \,\, l^a \,\, {\rm null,}
\end{eqnarray}
with $Q_a$ not dependent on $l^a$.
This is the equation which replaces (\ref{isoq-5.1}).
We get

\begin{eqnarray}\label{isoq-9.2}
2 \, \Big[(D-1) \, R_{ac} + \mu \, g_{ac}\Big] \, l^c =
Q_a, \ \ \ \forall \, l^a \, {\rm null},
\end{eqnarray}
which gives
$Q_a = 0$ and $Q_a l^a= 0$ identically
(for (\ref{isoq-9.2}) has to hold true e.g.
both for $l^c$ and $- l^c$), 
and then     
nothing more than case $(a)$.

Let us consider case $(c)$,
i.e. the case

\begin{eqnarray}\label{isoq-10.1}
Q_*(x, l^a) =
Q_{ab} l^a l^b \ne 0,  
\end{eqnarray}
with 
$Q_{ab}$ independent of $l^a$ and symmetric.
To require the vanishing of the variation of $F$ means to impose

\begin{eqnarray}\label{isoq-10.2}
\frac{\partial}{\partial l^a}
\bigg[(D-1) \, R_{cd} l^c l^d - Q_{cd} l^c l^d
+ \mu \, g_{cd} l^c l^d\bigg] = 0,
\,\,\,\, {\rm at} \,\, {\rm any} \,\, l^a \,\, {\rm null.}
\end{eqnarray}
From this we get

\begin{eqnarray}\label{isoq-10.3}
\Big[(D-1) \, R_{ac} - Q_{ac} + \mu \, g_{ac}\Big] \, l^c =
0, \ \ \ \forall \, l^a \, {\rm null}.
\end{eqnarray}
This implies

\begin{eqnarray}
\Big[(D-1) \, R_{ab} - Q_{ab} + \mu g_{ab}\Big] \, l^a l^b = 0, 
\ \ \ \forall \, l^a \, {\rm null},
\nonumber
\end{eqnarray}
which gives

\begin{eqnarray}\label{isoq-10.5}
\Big[(D-1) \, R_{ab} - Q_{ab}\Big] \, l^a l^b = 0, 
\ \ \ \forall \, l^a \, {\rm null}.
\end{eqnarray}

In case $(d)$,
$Q_*(x, l^a) = Q_{abc} l^a l^b l^c$,
with $Q_{abc}$ not dependent on $l^a$, and symmetric
in all its indices.
Starting from $F$ in (\ref{isoq-8.2}),
requiring irrelevance of $l^a$ means in this case

\begin{eqnarray}\label{isoq-12.2}
\frac{\partial}{\partial l^a}
\bigg[(D-1) \, R_{cd} l^c l^d - Q_{cde} l^c l^d l^e
+ \mu \, g_{cd} l^c l^d\bigg] = 0,
\,\,\,\, {\rm at} \,\, {\rm any} \,\, l^a \,\, {\rm null.}
\end{eqnarray}
This gives

\begin{eqnarray}\label{isoq-12.5}
2\, \Big[(D-1) \, R_{ac} + \mu \, g_{ac}\Big] \, l^c = 3 \, Q_{acd} l^c l^d,
\ \ \ \forall \, l^a \, {\rm null}.
\end{eqnarray}
Here, 
when sending $l^a$ in $-l^a$, the rhs does not change, while the lhs
flips the sign. We must then have ${\rm lhs} = {\rm rhs} = 0$,
that is $Q_{abc} l^c l^d = 0$. But this implies 
$Q_{acd} l^a l^c l^d = 0$, 
and we are back to case $(a)$.

Let us dispose now of case $(e)$.
Reconsidering what we just said for the case $Q_* = Q_{abc} l^a l^b l^c$,
we notice that for each further choice
$Q_* = Q_{abc ...} l^a l^b l^c ...$
with $r \equiv {\rm rank}(Q_{abc ...})$ odd,
we get an equation of the kind (\ref{isoq-12.5}),

\begin{eqnarray}
2\, \Big[(D-1) \, R_{ac} + \mu \, g_{ac}\Big] \, l^c = r \, Q_{acd} l^c l^d ...,
\ \ \ \forall \, l^a \, {\rm null}, \, {\rm with} \, \, r-1 \, \, \,
l^a{\rm s} \, \, {\rm in} \,\, {\rm the} \, \,  {\rm rhs},
\nonumber
\end{eqnarray} 
for which the same reasoning just described applies.
Then, the same as for the cases with $r=1$ and $r=3$,
all these further cases with $r$ odd 
turn out to give nothing more than case $(a)$.
When instead we take
$Q_* = Q_{abcd} l^a l^b l^c l^d$,
we have

\begin{eqnarray}\label{isoq-13.2}
\frac{\partial}{\partial l^a}
\bigg[(D-1) \, R_{cd} l^c l^d - Q_{cdef} l^c l^d l^e l^f
+ \mu \, g_{cd} l^c l^d\bigg] = 0,
\,\,\,\, {\rm at} \,\, {\rm any} \,\, l^a \,\, {\rm null.}
\end{eqnarray}
Following the by-now usual steps
we arrive at

\begin{eqnarray}\label{isoq-13.6}
\Big[(D-1) \, R_{ab} - 2 \, Q_{abcd} l^c l^d\Big] \, l^a l^b = 0, 
\ \ \ \forall \, l^a \, {\rm null}.
\nonumber
\end{eqnarray}
On the other hand,
(\ref{isoq-13.2}) directly implies

\begin{eqnarray}\label{isoq-13.7}
(D-1) \, R_{ab} l^a l^b = 
Q_{abcd} l^a l^b l^c l^d + \chi,
\ \ \ \forall \, l^a \, {\rm null}
\nonumber
\end{eqnarray}
with $\chi = \chi(x)$ a scalar not dependent on $l^a$.
Crossing the last two equations,
we get

\begin{eqnarray}\label{isoq-13.9}
Q_{abcd} l^a l^b l^c l^d = 
\chi,
\ \ \ \forall \, l^a \, {\rm null}.
\nonumber
\end{eqnarray}
But this is impossible,
unless 
$Q_{abcd} l^b l^c l^d = 0$ 
(from imposing the vanishing of the derivative of the lhs with respect
to $l^a$).
This however would give
$Q_{abcd} l^a l^b l^c l^d = 0$, and then again case $(a)$.
One can easily show that the same situation occurs
in any further even case, i.e. for $r>4$ even.

This concludes the discussion of the situation in which only one single
term with the $l^a$'s is present. 
One might wonder
that a generic combination of all these terms (the case we called $(f)$), 
could lead perhaps to something new
with respect to case $(c)$,
which we have seen is the only one able to add something
to case $(a)$ (including it), i.e. to what we considered in previous Section.
It is easily found however that this does not happen;
this is detailed in appendix \ref{AppB}. 
At the end,
what we have is thus that
case $(c)$, namely that with $r=2$, 
exhausts all possibilities
through which matter can act
to provide irrelevance of $l^a$ at $P$ after a probe.
As an aside,
we notice that
the Lagrangian multiplier in (\ref{isoq-10.2})
had no effect in (\ref{isoq-10.5})
(as well as the multiplier in (\ref{isoq-5.1})
had no effect in (\ref{isoq-5.5})).
This can be interpreted as suggesting
that in the analysis above
there is no need to restrict the variations of $l^a$
to give null $l'^a$'s ($l'^a \equiv l^a + \delta l^a$),
that is we can allow for variations to timelike or spacelike vectors $v'^a$
(with $l^a$ mapped continuously to $v'^a$). 

Armed with these findings,
we come back then to equation (\ref{isoq-10.5})
which sums up the results of case $(c)$. 
This equation shows that, 
thanks to matter acting as `generator' of curvature, we get
irrelevance of $l^a$ at $P$ after a probe,
with
$\lim_{\lambda \to 0} R_{(q)}(P, P') \ne 0$. 

The ordinary spacetime corresponding to this underlying minimum-length
spacetime is readily found in the same way 
(and the same maths \cite{PadN})
we followed in previous Section.
Equation (\ref{isoq-10.5}) implies

\begin{eqnarray}\label{isoq-10.7}
(D-1) \, R_{ab} - Q_{ab} =
\zeta \, g_{ab}
\end{eqnarray}
with $\zeta = \zeta(x)$ a scalar,
and this gives

\begin{eqnarray}\label{isoq-11.3}
G_{ab} =
\bigg(\frac{\zeta}{D-1} - \frac{1}{2} \, R\bigg) \, g_{ab} 
+ \frac{1}{D-1} \, Q_{ab}. 
\end{eqnarray}

This equation fixes a relation between the metric and matter source terms
being the latter expressed by tensor $Q_{ab}$. 
But, again, this is what are supposed to do the field equations.
We have thus the quite nice fact that
any ordinary spacetime obeying the (field) equations (\ref{isoq-11.3}) 
does admit a consistent qmetric description,
meaning a description in which the qmetric Ricci scalar
operationally expresses (according to the criterium we stated above)
the intrinsic geometry of spacetime,
as due.
 
If in equation (\ref{isoq-11.3})
we put $Q_{ab} = 0$,
we see that
from Bianchi identity and from the covariant constancy of $g_{ab}$
we get exactly equation (\ref{isoq-7.8});
this confirming that
what we called `empty space',
namely the case considered in the previous Section,
is indeed what we obtain using
the general equations in presence of matter, with matter removed.  

For $Q_{ab}$ generic,
from Bianchi identity we get

\begin{eqnarray}\label{isoq-11.4}
- \partial_b \bigg[\zeta - \frac{1}{2} (D-1) \, R\bigg] =
\nabla_a {Q^a}_b.
\end{eqnarray}
If $Q_{ab}$ is such that
(meaning, if the geometric scalar $Q$ associated to matter
is such that)

\begin{eqnarray}\label{isoq-11.5}
\nabla_a {Q^a}_b = 0,
\end{eqnarray} 
then

\begin{eqnarray}
\zeta - \frac{1}{2} (D-1) \, R = {\rm const},
\end{eqnarray}
and

\begin{eqnarray}\label{isoq-11.7}
G_{ab} =
C \, g_{ab} + \frac{1}{D-1} \, Q_{ab},
\end{eqnarray}
with $C$ the constant of (\ref{isoq-7.8}). 
If we take $D=4$ and $Q_{ab} = 24 \pi G T_{ab}$ (in units making
$c=1$ and $\hbar = 1$; $G$ is Newton constant),
we see these equations are Einstein' field equations with cosmological
constant,
implying this in particular that any spacetime
which is solution to full Einstein equations
does admit this consistent qmetric description. 
$Q_{ab}$ is what contains matter degrees of freedom;
it can in general depend also on the metric tensor,
on functions of it, on derivatives of arbitrary order,
as well as on additional fields.

%
%
%
%
%
%
%
%
%
%
\section{Discussion and conclusions}

What came out from the above,
is that the somehow puzzling aspect of the qmetric Ricci 
(bi)scalar $R_{(q)}$  
with base at a point $P$ of 
having a coincidence limit which depends
on the geodesic along which we reach $P$,
could actually be accommodated in a
spacetime which is given an operational meaning.
Specifically,
we discussed that
we ought to distinguish
between unprobed and probed spacetime at $P$,
the latter being the spacetime we get once a probe (of curvature)
at $P$ 
of original spacetime is done.
The assertion is that
$R_{(q)}$ at $P$ of unprobed spacetime is, in the small scale,
sort of multivalued function, or quantum superposition of different
potential values,
and the act of probing selects one of these.
Logical consistency demands then that a further probe
along any geodesic reaching $P$ on
the already-probed spacetime gives that same value
obtained in the first probe.
What has been shown happens afterwards,
is that this requirement
goes hand in hand with regarding matter as capable to affect 
large-scale geometry. 

If the geometric contents of matter are incorporated
in a geometric scalar $Q$ 
(much the same way as the Ricci scalar $R$ does for geometry itself),
with minumum-length counterpart $Q_{(q)}$, 
the only case relevant for obtaining independence from 
the tangent $l^a$ to the geodesic at $P$
after a probe
is that
which gives 
$\lim_{L_0 \to 0} \, \lim_{p \to P} \, Q_{(q)}(p, P) \equiv
Q_* =
Q_{ab}(x) l^a l^b$,
i.e. the limit has quadratic dependence on $l^a$,
and the
$l^a$-independence translates
into relations (\ref{isoq-11.3}) for ordinary metric,
which are field equations.
A specific choice of $Q_{ab}$ exhibits
equation (\ref{isoq-11.3}) as the Einstein field equations
with cosmological constant,
this shows that any spacetime solution to these equations
does admit a consistent (in the operational sense above) qmetric description.

Reconsidering the route we have followed,
things go also on the reverse. 
If
a spacetime endowed with a limit length $L_0$,
does admit an operationally consistent metric-like description
in the small scale,
then in the large scale (i.e. where it goes to coincide
with an ordinary spacetime)
it obeys field equations.
Thus,
if the spacetime we have got to describe
has actually a limit length,
this implies that in the large scale
this spacetime
has to obey field equations.
This resonates 
with what expressed 
in \cite{JacB} complemented with \cite{JacD}   
(cf. also \cite{PesG, PesH};
in present case
however in terms
of a larger class of possible field equations).

What
the (large scale) field equations
turn out to be connected with, 
is the requirement that
a generic spacetime endowed with a (lower) limit length
(on which origin we know nothing apart assuming that its existence involves 
quantum effects),
does have an operationally-meaningful expression for curvature,
sort thus this of logical consistency.
From the expression of the scalar $F$
(equations (\ref{isoq-8.2}) and (\ref{isoq-10.1})), namely of the quantity
that must be $l^a$-independent to have consistency,
we noticed that
the vanishing of the variation
finds interpretation as a
request of balance in the exchange of heat associated
to spacetime degrees of freedom ($(1/L_{Pl}^2) \, R_{cd} l^c l^d$)
and heat associated 
to matter degrees of freedom ($(1/L_{Pl}^2) \, Q_{cd} l^c l^d$).
This is
exactly the same condition, with the same maths, 
which is present in the derivation \cite{PadG_, PadF, Pad08, Pad20}
of Einstein full field equations
from a thermodynamic variational principle,
consisting the latter in requiring the balance of exchanged heats.
What we find here then
is that the very physical principle
of requiring the balance of exchanged heats
is nothing else than the expression of
logical consistency of a spacetime endowed 
with the operational meaning described above.
Moreover,
in a spacetime  
actually endowed with a limit length,
not only this spacetime has to obey field equations in the large scale,
but it is this physical principle (of balance of exchanged heats)
which,
on pain of operational incongruences, 
logically requires them.
 
Many have been the attempts which dreamed
of the existence of a thermodynamic principle,
conceived as more fundamental than field equations themselves,
from which the latter could be derived.
This study somehow adds to them
bolstering the request of balance of exchanged heats
as the thermodynamic principle sought-after.
Moreover,
this thermodynamic principle 
is pointed out to be connected 
with a specific requirement of a consistent
operational description of spacetime
from the large down to the smallest conceivable scale.

The superposition of values of $R_{(q)}$
for unprobed spacetime,
is an effect of a $L_0 \ne 0$.
The thing is that
this feature
keeps remaining also in the $L_0 \to 0$ limit.
We are then confronted with two scenarios
which do result inequivalent:
i) absence of any limit length 
($l^a$-independence at $P$ obvious,
for the coincidence limit would be $R$; 
no requirement of large-scale
field equations);
ii) presence of a vanishingly-small limit length
($l^a$-independence after a probe not obvious; 
requirement of large-scale field equations).
But, large-scale field equations do exist 
for actual spacetime (Einstein field equations,
to the best of experimental checks),
and they indeed foresee a limit length when combined
with basic principles of quantum mechanics.
This selects scenario (ii), 
and, at the same time, indicates 
that the request of existence of large-scale field equations by this scenario
is insensitive to the actual (provided very small) value of $L_0$.
This suggests that field equations,
and in particular Einstein field equations,
ought to be regarded as quantum in their origin,
even if at conditions at which
the quantum nature of spacetime can hardly be directly probed
by effects small with $L_0$
(this adding to what suggested in \cite{PadB_3}).
Thanks to the persistence of large-scale quantum effects
also in the $L_0 \to 0$ limit,
are the Einstein equations themselves 
what testifies about spacetime being quantum. 
In view of this,
we might consider the field equations
as `zero-point' field equations,
with the meaning of something
which quantum-mechanically stays there
while classically it would not.

Then,
the field equations ought not to be considered
as the classical limit of a quantum theory of gravity
(meaning the equations we would obtain when letting
$L_0 \to 0$ with $\hbar \to 0$),
but a direct prediction of this quantum theory.
In other words,
we do not get the large-scale field equations
(e.g. the Einstein equations)  in the $\hbar\to 0$ limit;
rather, the large-scale field equations arise,
find their origin, in an explicitly $L_0 \ne 0$ (and then $\hbar \ne 0$),
and stay there even when $L_0$ becomes exceedingly and unappreciably small.
They do not `set in' in the $L_0 \to 0$ limit; 
on the contrary, what they do is to keep staying there also in this
limit.
They are sort of quantum effect not vanishing with~$\hbar$.

As a closing remark,
we would like to emphasize
that all this is not about what the field equations ought
to become,
or what they ought to be replaced by,
in the small scale.
No word is told about that up to this point in the paper.
Everything we described, is only about the connections
that the endowing of spacetime with a (lower) limit length 
seemingly turns out to have
with {\it large scale} (as compared with Planck length) structure.
Clearly, coping with the short scale,
would imply to have to do with a length scale
at which we can no longer reasonably neglect the own evolution
of the dofs
of the specific microscopic theory, 
as we do instead in our approximation.
If we apply our model anyway,
we notice that
when the scale is short enough that
$g_{ab}$ is no longer a good approximation of the qmetric,
the constraints (Eq. (\ref{isoq-11.3}))
keep remaining formally the same,
but what they constrain (i.e. $g_{ab}$) has no longer 
the meaning which we operationally assign to a metric
(i.e. to give quadratic intervals), and for this we should
refer instead to the qmetric.
Inverting in these equations from $g_{ab}$ to the qmetric, 
would give the evolution equations in the short scale
for the (effective) metric.

{\it Acknowledgements}.
I am grateful to Francesco Anselmo for drawing attention to one of the
references. I would like also to thank Sumanta Chakraborty and
Dawood Kothawala for consideration of a draft of the paper.

\appendix

%
%
%
%
%
%
%
\section{Derivation of equalities (\ref{isoq-20.1})}\label{AppA}

The quantity
$\lim_{L_0 \to 0} \lim_{p \to P} R_{(q)}(p, P)$ has been already
evaluated in \cite{PesP}. 
What we add here,
is an expression for  $\lim_{p \to P} R_{(q)}(p, P)$
at finite $L_0$,
detailing,  in a conveniently chosen parameter, 
the order of magnitude
of the non-leading part,
this way justifying expressions (\ref{isoq-20.1}). 

To this aim,
we start from equation (\ref{Rq}),   
rewritten in a convenient, slightly modified form
(corresponding to merge two of its terms into one,
and leave the others unchanged):

\begin{eqnarray}\label{isoq-37.1}
R_{(q)}(p, P) &=& 
\frac{1}{A} \, R_\Sigma
- 2 \, \alpha \, \frac{d\alpha}{d\lambda} \, K
+ 2 \, \alpha^2 \, R_{ab} \, l^a l^b
- (D-2) \, \alpha \, \frac{d}{d\lambda}
\Big(\alpha \, \frac{d}{d\lambda} \ln A\Big)
\nonumber \\
& &
- \frac{1}{4} (D-2) (D-1) \, \alpha^2 \, 
\Big(\frac{d}{d\lambda} \ln A\Big)^2
- \alpha^2 K^2
+ \alpha^2 K^{ab} K_{ab}
- (D-1) \, \alpha^2 \,
\Big(\frac{d}{d\lambda} \ln A\Big) K.
\end{eqnarray}

In this formula,
first we provide expressions of
the (four) terms not containing $A$.
They are (\cite{PesP}, cf. \cite{KotG, KotI})

\begin{eqnarray}
\label{isoq-29.1}
- 2 \, \alpha \, \frac{d\alpha}{d\lambda} \, K 
&=&
- 2 \, \alpha \, \frac{d\alpha}{d\lambda} \,
\frac{D-2}{\lambda} + {\cal O}(\lambda),
\\
\label{isoq-29.3}
2\, \alpha^2 \, R_{ab} l^a l^b 
&=&
2\, \alpha^2 \, E(p),
\\
\label{isoq-29.4}
- \alpha^2 K^2 
&=&
- (D-2)^2 \, \alpha^2 \, \frac{1}{\lambda^2}
+ \frac{2}{3} (D-2) \, \alpha^2 \, E(p) + {\cal O}(\lambda),
\\
\label{isoq-29.5}
\alpha^2 \, K^{ab} K_{ab}
&=&
(D-2) \, \alpha^2 \, \frac{1}{\lambda^2} 
- \frac{2}{3} \, \alpha^2 E(p) + {\cal O}(\lambda).
\end{eqnarray}
with, even where not explicitly indicated,
all quantities evaluated at $p$
and \cite{Pad01}
$E_{ab} \equiv R_{ambn} l^m l^n$
with
$E(p) \equiv {E^a}_a(p) = (R_{ab} l^a l^b)(p)
\ge 0$, with the last relation for
our spacetime obeys the null convergence condition.
Here, we used the expression

\begin{eqnarray}
K_{ab} = \frac{1}{\lambda} \, h_{ab}
- \frac{1}{3} \, \lambda \, {h^c}_a {h^d}_b E_{cd} + {\cal O}(\lambda^2),
\end{eqnarray}
and thus also

\begin{eqnarray}
K = 
(D-2) \, \frac{1}{\lambda} - \frac{1}{3} \, \lambda E(p) +
{\cal O}(\lambda^2),
\end{eqnarray}
from \cite{PesP} (but cf. \cite{KotG, KotI}).
Expressions (\ref{isoq-29.1}), (\ref{isoq-29.4}) and (\ref{isoq-29.5})
have, at leading order in $\lambda$,
factors $1/\lambda$ or $1/\lambda^2$ in them, divergent
in the $p \to P$ (i.e. $\lambda \to 0$) limit.
The actual divergence or not of the whole expressions
would depend also on the behavior $\alpha$ and $d\alpha/d\lambda$
in the same limit.
A divergent $\alpha$ would also introduce a divergence in the
remaining term $(\ref{isoq-29.3})$,
and also e.g. the ${\cal O}(\lambda)$ term in (\ref{isoq-29.1})
could be diverging in the $\lambda \to 0$ limit.
As we will see,
it turns out however that the remaining terms in the expression 
of $R_{(q)}(p, P)$
do cancel any $\lambda^{-2}$, $\lambda^{-1}$, $\lambda^0$ term here,
i.e. it turns out they do not give any contribution to $R_{(q)}(p, P)$
whichever is the assumed behaviour of $\alpha$ and $d\alpha/d\lambda$
in the $\lambda \to 0$ limit.
It remains an open question however what happens to the ${\cal O}(\lambda)$
terms (do the cancellations extend to these, and higher, orders?).
To handle this,
we assume then that
$\alpha$ and $d\alpha/d\lambda$ remain finite
in the $\lambda \to 0$ limit.
This guarantees that 
each ${\cal O}(\lambda)$ term vanishes with $\lambda$
(as well as any higher order term).
Strictly speaking, the derivation we are providing here is thus
for non-divergent $\alpha$ and $d\alpha/d\lambda$. 

As for the terms containing $A$,
from the expression (\ref{A}) for it
we see that what we need is the expansion
around $P$ of the van Vleck determinant,
which, for $E(p)$ smooth at $P$, in our circumstances 
reads \cite{Xen}

\begin{eqnarray}\label{isoq-26.2}
\Delta^{1/2}(p, P) =
1 + \frac{1}{12} \, \lambda^2 E(p)
- \frac{1}{24} \, \lambda^3 \frac{dE}{d\lambda}(p)
+ \lambda^4 
\bigg[\frac{1}{288} \, E^2(p)
+ \frac{1}{360} \, E^{ab}(p) E_{ab}(p) 
+ \frac{1}{80} \, \frac{d^2E}{d\lambda^2}\bigg]
+ {\cal O}(\lambda ^5),
\end{eqnarray}
from which

\begin{eqnarray}\label{isoq-26.4}
{\tilde \Delta}^{1/2}(p, P) =
1 + \frac{1}{12} \, \tilde\lambda^2 E(\tilde p)
- \frac{1}{24} \, \tilde\lambda^3 \frac{dE}{d\tilde\lambda}({\tilde p})
+ \tilde\lambda^4 
\bigg[\frac{1}{288} \, E^2({\tilde p})
+ \frac{1}{360} \, E^{ab}({\tilde p}) E_{ab}({\tilde p}) 
+ \frac{1}{80} \, \frac{d^2E}{d\lambda^2}(\tilde p)\bigg]
+ {\cal O}(\tilde\lambda^5),
\end{eqnarray}
which is $\Delta^{1/2}$ evaluated at $\tilde p$ along $\gamma$, with 
$\lambda(\tilde p, P) = \tilde\lambda$.

The very writing of these expansions
comes with the desire
that it can happen
that any term at an assigned order in powers 
of $\lambda$ or in $\tilde\lambda$
turns out generically negligible 
with respect to that at the previous order.
Our first task is to try to characterize this,
in the sense of finding a convenient parameter,
for which we can say if, at coincidence limit, 
it is small enough 
to give what just said.
This parameter can be clearly $\lambda$ itself
for the expansion (\ref{isoq-26.2}),
for in the coincidence limit it becomes vanishingly small.
It appears perhaps not so clear 
what happens instead for expansion (\ref{isoq-26.4}).
Here, $\tilde\lambda$ remains finite at coincidence,
and we need some reference scale to compare with
to establish if $\tilde\lambda$ becomes actually small enough
to provide full meaning to the expansion.  
Let us focus then on expansion (\ref{isoq-26.4}). 
 
First of all,
from the $\tilde\lambda^2$-term we see that
we need to be at conditions in which
$\tilde\lambda^2 E(\tilde p) \ll 1$. 
To characterize this,
for the given null geodesic $\gamma$ with tangent 
$l^a = dx^a/d\lambda$, 
we introduce then a scale length $\ell_R$ associated 
to curvature (of the assigned spacetime) at any given event 
(near $P$)
as
$\ell_R \equiv 1/\sqrt{E}$ at that event
($E$ non-negative, for null convergence condition assumed to hold).
We have to ask that curvature is small,
and clearly it is small enough 
if we have
$\tilde\lambda/\ell_R \simeq L_0/\ell_R \ll 1$,
where 1st relation comes from assuming
to be at conditions $\tilde\lambda \simeq L_0$,
i.e. to be near the coincidence limit ($\lambda \simeq 0$).
This gives indeed 
$
\tilde\lambda^2 E(\tilde p) =
{\cal O}(\tilde\lambda^2/\ell_R^2)
\ll 1.
$
As mentioned in the main text,
this corresponds to that, 
in a local frame in which $\lambda$ is length,
$\bar p$ at $\lambda(\bar p, P) = L_0$
results near enough to $P$ to give
$|g_{ab}(\bar p)| = {\cal O}(R_{abcd}) \, L_0^2 \ll 1$
(using Riemann normal coordinates).

Next,
let us write

\begin{eqnarray}\label{isoq-27.5}
\ell_R(\tilde p) =
L_R + C \, \lambda(\tilde p, P) 
+ C_2 \, \frac{\lambda^2(\tilde p, P)}{L_R}
+ C_3 \, \frac{\lambda^3(\tilde p, P)}{L_R^2} + ... \, ,
\end{eqnarray}
with 
$L_R \equiv 1/\sqrt{E(P)}$,
and $C$, $C_1$, $C_2$, .. constants, 
and where we have put in evidence as much $1/L_R$ 
factors as dimensionally required.  
This expansion shows that $\tilde\lambda/L_R$
is a parameter
which is indeed effective in discriminating how significantly
$\ell_R$ differs from its value $L_R$ at $P$.
Having this,
we consider the next term, the $\tilde\lambda^3$-term, 
in (\ref{isoq-26.4}). 
We have

\begin{eqnarray}\label{isoq-27.4}
\frac{dE}{d\tilde\lambda}(\tilde p) 
&=&
\frac{dE}{d\ell_R} \, \frac{d\ell_R}{d\tilde\lambda}
\nonumber \\
&=&
- C \, \frac{1}{\ell_R^3}
\nonumber \\
&=&
- C \, \frac{1}{\ell_R} \, E(\tilde p)
\nonumber \\
&=&
- C \, \frac{1}{L_R} \, E(\tilde p)
+ {\cal O}\bigg(\frac{\tilde\lambda}{L_R} \, \frac{E(\tilde p)}{L_R}\bigg),
\end{eqnarray}
where use has been made of (\ref{isoq-27.5}).
This gives
$
\tilde\lambda \, \frac{dE}{d\tilde\lambda}(\tilde p) =
{\cal O}\Big(\frac{\tilde\lambda}{L_R} \, E(\tilde p)\Big)
$
and
$
\tilde\lambda^3 \frac{dE}{d\tilde\lambda}(\tilde p) =
\tilde\lambda^2{\cal O}\Big(\frac{\tilde\lambda}{L_R} \, E(\tilde p)\Big).
$

In the $\tilde\lambda^4$-term,
we have
$
\tilde\lambda^4 \, E^2(\tilde p) =
\tilde\lambda^2 E(\tilde p) \, \frac{\tilde\lambda^2}{\ell_R^2} =
\tilde\lambda^2 E(\tilde p) \, \frac{\tilde\lambda^2}{L_R^2} \,
\Big(1 + {\cal O}\big(\frac{\tilde\lambda}{L_R}\big)\Big) =
\tilde\lambda^2 {\cal O}\Big(\frac{\tilde\lambda^2}{L_R^2} \, E(\tilde p)\Big);
$
further,
$
\tilde\lambda^4 \, E^{ab}(\tilde p) E_{ab}(\tilde p) =
{\cal O}\big(\tilde\lambda^4 E^2(\tilde p)\big) =
\tilde\lambda^2 {\cal O}\Big(\frac{\tilde\lambda^2}{L_R^2} \, E(\tilde p)\Big)
$
too (from evaluating the scalar $E^{ab} E_{ab}$ in the local frame
in which $\lambda$ is length);
and
\begin{eqnarray}
\frac{d^2E}{d\tilde\lambda^2}(\tilde p) 
&=&
\frac{d}{d\tilde\lambda}
\bigg(- \frac{1}{\ell_R^3} \, \frac{d\ell_R}{d\tilde\lambda}\bigg)(\tilde p)
\nonumber \\
&=&
\bigg(3 \, \frac{1}{\ell_R^4} \, C 
- \frac{1}{\ell_R^3} \, \frac{2 \, C_2}{L_R}\bigg)(\tilde p)
\nonumber \\
&=&
\frac{1}{L_R^2} \, E(\tilde p)
\, \big(3 \, C - 2 \, C_2\big) 
\bigg(1 + {\cal O}\Big(\frac{\tilde\lambda}{L_R}\Big)\bigg)
\nonumber \\
&=&
{\cal O}\bigg(\frac{1}{L_R^2} \, E(\tilde p)\bigg).
\nonumber
\end{eqnarray}
All this, gives 
$\tilde\lambda^2 {\cal O}\Big(\frac{\tilde\lambda^2}{L_R^2} 
\, E(\tilde p)\Big)$
as order of magnitude of the whole $\tilde\lambda^4$-term.  
We can proceed in a similar manner at any order in $\tilde\lambda$.
In each $\tilde\lambda^n$ term, there will be factors of powers 
$(E_{ab})^m$ or $E^m$, derivatives $d^{m'}E/d\tilde\lambda^{m'}$, 
with $m$, $m'$ integers $\ge 0$ such that 
$2 \, m + m' + 2 = n$, as dimensionally required.    
And this implies that
the $n$-th order term will be
$
{\cal O}\Big(\frac{\tilde\lambda^n}{L^n}\Big)
= \tilde\lambda^2 
{\cal O}\Big(\frac{\tilde\lambda^{n-2}}{L_R^{n-2}} \, E(\tilde p)\Big).
$

Summing all up,
we can rewrite (\ref{isoq-26.4}) as

\begin{eqnarray}
\label{isoq-28.8}
{\tilde \Delta}^{1/2}(p, P) 
&=&
1 + \frac{1}{12} \, \tilde\lambda^2 E(\tilde p)
- \frac{1}{24} \, \tilde\lambda^3 \, \frac{dE}{d\tilde\lambda}(\tilde p)
+ \tilde\lambda^2 
{\cal O}\bigg(\frac{\tilde\lambda^2}{L_R^2} \, E(\tilde p)\bigg)
\\
&=&
\label{isoq-28.8_bis}
1 + \frac{1}{12} \, \tilde\lambda^2 E(\tilde p)
+ \tilde\lambda^2 
{\cal O}\bigg(\frac{\tilde\lambda}{L_R} \, E(\tilde p)\bigg)
+ \tilde\lambda^2 
{\cal O}\bigg(\frac{\tilde\lambda^2}{L_R^2} \, E(\tilde p)\bigg),
\end{eqnarray}
where in  
(\ref{isoq-28.8_bis}) 
we explicitly write the order of magnitude
of the 2nd term in the rhs of (\ref{isoq-28.8}).

We can proceed now to compute the expressions of the terms
containing $A$ in (\ref{isoq-37.1}) in the coincidence limit.
For the 1st term, we get

\begin{eqnarray}\label{isoq-31.5}
\frac{1}{A} \, R_{\Sigma} 
&=&
\lambda^2 \, \Delta^{-\frac{2}{D-2}} 
\Big(R(p) + K^2(p) -K^{ab}(p) K_{ab}(p) - 2 \, E(p)\Big)
\, \frac{1}{\tilde\lambda^2}
\, \tilde \Delta^{\frac{2}{D-2}}
\nonumber \\
&=&
\Big((D-2)(D-3) + {\cal O}(\lambda^2)\Big)
\, \frac{1}{\tilde\lambda^2} \,
\bigg(1 + \frac{1}{3 (D-2)} \, \tilde\lambda^2 \, E(\tilde p)
+ \tilde\lambda^2 
{\cal O}\Big(\frac{\tilde\lambda}{L_R} \, E(\tilde p)\Big)\bigg)
\nonumber \\
&=&
(D-2)(D-3) \, \frac{1}{\tilde\lambda^2}
+ \frac{D-3}{3} \, E(\tilde p) 
+ {\cal O}\Big(\frac{\tilde\lambda}{L_R} \, E(\tilde p)\Big)
+ {\cal O}(\lambda^2), 
\end{eqnarray}
where we used of relation (\ref{isoq-2.2}) (2nd equality), of the expressions
(\ref{isoq-29.4}) and (\ref{isoq-29.5}) for $K^2$ and $K^{ab} K_{ab}$,
as well as of the expansions of the van Vleck determinant.

As for the 4th term in (\ref{isoq-37.1}),
we notice first that

\begin{eqnarray}\label{isoq-32.3}
\alpha \, \frac{d}{d\lambda} \ln A 
&=&
\frac{2}{\tilde\lambda}
- 2 \, \alpha \, \frac{1}{\lambda}
- \alpha \, \frac{d}{d\lambda}
\bigg[\frac{1}{3 (D-2)} \, \tilde\lambda^2 \, E(\tilde p)
- \frac{1}{6 (D-2)} \, \tilde\lambda^3 \, \frac{dE}{d\tilde\lambda}(\tilde p)
+ {\cal O}\Big(\frac{\tilde\lambda^2}{L_R^2} 
\, \tilde\lambda^2 E(\tilde p)\Big)\bigg]
\nonumber \\
& &
- \alpha \, \frac{d}{d\lambda}
\bigg[- \frac{1}{3 (D-2)} \, \lambda^2 \, E(p) + {\cal O}(\lambda^3)\bigg]
\nonumber \\
&=&
\frac{2}{\tilde\lambda} 
- \frac{2}{3 (D-2)} \, \tilde\lambda \, E(\tilde p)
+ \frac{1}{6 (D-2)} \, \tilde\lambda^2 \, \frac{dE}{d\tilde\lambda}(\tilde p)
+ {\cal O}\Big(\frac{\tilde\lambda^2}{L_R^2} \, \tilde\lambda E(\tilde p)\Big)
\nonumber \\
& &
- 2\, \alpha \, \frac{1}{\lambda}
+ \frac{2}{3 (D-2)} \, \alpha \, \lambda \, E(p) 
+ {\cal O}(\lambda^2),
\end{eqnarray}
where 
$
\tilde\lambda^2 \frac{dE}{d\tilde\lambda}(\tilde p) =
{\cal O}\big(\frac{\tilde\lambda}{L_R} \, \tilde\lambda E(\tilde p)\big)
$
and we used of
$
\frac{d}{d\lambda} {\cal O}\big(\frac{\tilde\lambda^3}{L_R} E(\tilde p)\big)
= \frac{1}{\alpha} \big[{\cal O}\big(\frac{\tilde\lambda}{L_R} 
\, \tilde\lambda E(\tilde p)\big) +
{\cal O}\big(\frac{\tilde\lambda^2}{L_R^2} 
\, \tilde\lambda E(\tilde p)\big)\big]
$
and similarly at any order.
This then gives

\begin{eqnarray}\label{isoq-34.4}
- (D-2) \, \alpha \, \frac{d}{d\lambda}
\Big(\alpha \, \frac{d}{d\lambda} \ln A\Big) 
&=&
2 (D-2) \, \frac{1}{\tilde\lambda^2}
+ \frac{2}{3} \, E(\tilde p)
+ {\cal O}\Big(\frac{\tilde\lambda}{L_R} \, E(\tilde p)\Big)
\nonumber \\
& &
- 2 (D-2) \, \alpha^2 \, \frac{1}{\lambda^2}
+ 2 (D-2) \, \alpha \, \frac{d\alpha}{d\lambda} \, \frac{1}{\lambda}
- \frac{2}{3} \, \alpha^2 \, E(p) 
+ {\cal O}(\lambda).
\end{eqnarray}

As for the remaining two terms in (\ref{isoq-37.1}),
namely the 5th and the 8th, there is a convenience
in treating them together.
In fact, we have

\begin{eqnarray}\label{isoq-36.5}
& & - \frac{1}{4} (D-2) (D-1) \, \alpha^2 \, 
\Big(\frac{d}{d\lambda} \ln A\Big)^2
- (D-1) \, \alpha^2 \,
\Big(\frac{d}{d\lambda} \ln A\Big) K 
\nonumber \\
&=&
- \frac{1}{4} (D-2)(D-1) \,
\Big(\alpha \, \frac{d}{d\lambda} \ln A\Big) \times
\nonumber \\
& &
\bigg[\frac{2}{\tilde\lambda} 
- \frac{2}{3 (D-2)} \, \tilde\lambda \, E(\tilde p)
+ \frac{1}{6 (D-2)} \, \tilde\lambda^2 \, \frac{dE}{d\tilde\lambda}(\tilde p)
+ {\cal O}\Big(\frac{\tilde\lambda^2}{L_R^2} 
\, \tilde\lambda E(\tilde p)\Big)
- 2\, \alpha \, \frac{1}{\lambda}
+ \frac{2}{3 (D-2)} \, \alpha \, \lambda \, E(p) 
+ {\cal O}(\lambda^2)\bigg]
\nonumber \\
& &
- (D-1) \, \alpha \, \Big(\alpha \, \frac{d}{d\lambda} \ln A\Big) \,
\Big((D-2) \, \frac{1}{\lambda} - \frac{1}{3} \, \lambda \, E(p)
+ {\cal O}(\lambda^2)\Big) 
\nonumber \\
&=&
\frac{1}{2} (D-2)(D-1) \, \alpha^2 \, \frac{1}{\lambda} \, 
\frac{d}{d\lambda} \ln A 
- \frac{1}{4} (D-2)(D-1) \,
\bigg[ ...\bigg] \,
\bigg\{\bigg[ ...\bigg] + 2 \, \alpha \, \frac{1}{\lambda}\bigg\}
\nonumber \\
& &
- (D-2)(D-1) \, \alpha^2 \, \frac{1}{\lambda} \,
\frac{d}{d\lambda} \ln A 
+ \frac{D-1}{3} \, \alpha \, \lambda \, E(p) \, \bigg[ ...\bigg] 
- (D-1) \, \alpha \, {\cal O}(\lambda^2) \, \bigg[ ...\bigg]
\nonumber \\
&=&
- \frac{1}{4} (D-2)(D-1) \, \bigg\{ ...\bigg\}^2
+ (D-2)(D-1) \, \alpha^2 \, \frac{1}{\lambda^2}
- \frac{2}{3} (D-1) \, \alpha^2 \, E(p) 
+ {\cal O}(\lambda)
\nonumber \\
&=&
- (D-2)(D-1) \, \frac{1}{\tilde\lambda^2}
+ \frac{2}{3} (D-1) \, E(\tilde p)
- \frac{D-1}{6} \, \tilde\lambda \, \frac{dE}{d\tilde\lambda}(\tilde p)
+ {\cal O}\Big(\frac{\tilde\lambda^2}{L_R^2} \, E(\tilde p)\Big)
\nonumber \\
& &
+ (D-2)(D-1) \, \alpha^2 \, \frac{1}{\lambda^2}
- \frac{2}{3} (D-1) \, \alpha^2 \, E(p)
+ {\cal O}(\lambda)
\end{eqnarray}
(where $\big[ ...\big]$ stands for $\alpha \frac{d}{d\lambda} \ln A$ 
expanded, i.e. what is written in square brackets in the 1st equality,
and $\big\{ ...\big\}$ denotes the quantity in braces in the 2nd equality),
and in 3rd equality we see that the 
$\frac{1}{\lambda} \frac{d}{d\lambda}\ln A$ terms
nicely cancel.
The quantity $\tilde\lambda \, \frac{dE}{d\tilde\lambda}(\tilde p)$,
we know is ${\cal O}\big(\frac{\tilde\lambda}{L_R} \, E(\tilde p)\big)$.

Putting all this together,
i.e. substituting equations
(\ref{isoq-29.1}-\ref{isoq-29.5})
and (\ref{isoq-31.5}), (\ref{isoq-34.4}), (\ref{isoq-36.5})
into equation (\ref{isoq-37.1}),
we finally get

\begin{eqnarray}
R_{(q)}(p, P) = 
(D-1) \, E(\tilde p) 
+ {\cal O}\Big(\frac{\tilde\lambda}{L_R} \, E(\tilde p)\Big)
+ {\cal O}(\lambda).
\end{eqnarray}
Then,

\begin{eqnarray}
\label{isoq-38.1}
\lim_{p \to P} R_{(q)}(p, P) 
&=&
(D-1) \, E(\bar p)
+ {\cal O}\Big(\frac{L_0}{L_R} \, E(\bar p)\Big)
\\
&=&
(D-1) \,
\bigg(E(P) + E(P) \, {\cal O}\Big(\frac{L_0}{L_R}\Big)\bigg) 
+ {\cal O}\Big(\frac{L_0}{L_R} \, E(\bar p)\Big)
\nonumber \\
\label{isoq-38.4}
&=&
(D-1) \, E(P)
+ {\cal O}\Big(\frac{L_0}{L_R} \, E(P)\Big),
\end{eqnarray}
where $\bar p$ is such that $\lambda(\bar p, P) = L_0$,
and we used
$
E(\tilde p) = E(P) + E(P) \, {\cal O}\big(\frac{\tilde\lambda}{L_R}\big),
$
which gives
$
\frac{\tilde\lambda}{L_R} \, E(\tilde p) =
\frac{\tilde\lambda}{L_R} \, E(P) 
+ {\cal O}\big(\frac{\tilde\lambda^2}{L_R^2} \, E(P)\big).
$
(\ref{isoq-38.1}) and (\ref{isoq-38.4}) are the
equalities (\ref{isoq-20.1}) of the main text.

%
%
%
%
%
%
%
\section{Consideration of case $(f)$ (see text)}\label{AppB}


For the quantity $Q_*$,
i.e. the small scale limit of the qmetric quantity which captures the
geometrical effects of matter,
we already considered in the main text
all the cases in which
$Q_*$ has no dependence on $l^a$, a linear dependence, a quadratic,
a cubic, .. , separately.
Our task here is to establish whether the case of a generic combination 
of all the cases above adds something or not.

To this aim,
let us write

\begin{eqnarray}\label{isoq-15.1}
Q_* =
\sum_{r = 0}^n \,
Q_{12 ... r} \, l^1 l^2 ... l^r,
\end{eqnarray}
where indices $1, 2, ... , r$ are short for indices
$a_1, a_2, ..., a_r$ with each $a_i = 1, ..., D$.
Tensors $Q_{12 ... r}$ do not depend on $l^a$.
They can be taken
totally symmetric without
loss of generality.
Moreover, we think of any coefficient 
(necessarily independent from $l^a$)
in the linear combination
(\ref{isoq-15.1}) 
as absorbed into $Q_{12 ... r}$ themselves. 

From $F$ as in (\ref{isoq-8.2}),
we get path-independence at the path assigned 
if we require 

\begin{eqnarray}\label{isoq-15.2}
\frac{\partial}{\partial l^a}
\bigg[
(D-1) \, R_{cd} l^c l^d 
- \sum_{r = 0}^n Q_{12 ... r} \, l^1 l^2 ... l^r
- \mu \, l^c l_c
\bigg],
\end{eqnarray}
with $\mu$ a scalar not dependent on $l^a$.
We get

\begin{eqnarray}
\nonumber
2 (D-1) \, R_{ac} l^c
- \sum_{r = 1}^n \, r \, Q_{a12 ... r-1} \, l^1 l^2 ... l^{r-1}
- 2 \, \mu \, l_a
= 0, \; \; \; \; \; \; \; \; \forall l^a \, {\rm null},
\end{eqnarray}
whence

\begin{eqnarray}
\nonumber
2 (D-1) \, R_{ac} l^c
- 2 \, \mu \, l_a
- {\sum_{r = 2}^n}_{\, \,({\rm only} \, \, r \, \, {\rm even})}  \, r \, 
Q_{a12 ... r-1} \, l^1 l^2 ... l^{r-1}
=
{\sum_{r = 1}^n}_{\, \,({\rm only} \, \, r \, \, {\rm odd})}  \, r \, 
Q_{a12 ... r-1} \, l^1 l^2 ... l^{r-1},
\; \; \; \; \; \; \; \; \forall l^a \, {\rm null}.
\end{eqnarray}
Here we see that sending $l^a$ in $-l^a$ the lhs changes sign
while the rhs does not.
This implies 
${\rm lhs} = 0 = {\rm rhs} \, \, \, \forall l^a$, 
then
we cannot have terms with $r$ odd in sum (\ref{isoq-15.1}).

We are left with

\begin{eqnarray}\label{isoq-16.1}
2 (D-1) \, R_{ac} l^c
- 2 \, \mu \, l_a
- {\sum_{r = 2}^n}_{\, \,({\rm only} \, \, r \, \, {\rm even})}  \, r \, 
Q_{a12 ... r-1} \, l^1 l^2 ... l^{r-1}
=
0, \; \; \; \; \; \; \; \; \forall l^a \, {\rm null}
\end{eqnarray}
which can be rewritten as

\begin{eqnarray}
\nonumber
\bigg[
2 (D-1) \, R_{ac}
- 2 \, \mu \, g_{ac}
- {\sum_{r = 2}^n}_{\, \,({\rm only} \, \, r \, \, {\rm even})}  \, r \, 
Q_{ac12 ... r-2} \, l^1 l^2 ... l^{r-2}
\bigg] \, l^c = 0.
\; \; \; \; \; \; \; \; \forall l^a \, {\rm null}
\end{eqnarray}
If this is true, it is true also

\begin{eqnarray}
\nonumber
\bigg[
2 (D-1) \, R_{ab}
- 2 \, \mu \, g_{ab}
- {\sum_{r = 2}^n}_{\, \,({\rm only} \, \, r \, \, {\rm even})}  \, r \, 
Q_{ab12 ... r-2} \, l^1 l^2 ... l^{r-2}
\bigg] \, l^a l^b = 0,
\; \; \; \; \; \; \; \; \forall l^a \, {\rm null},
\end{eqnarray}
which is

\begin{eqnarray}\label{isoq-16.4}
\bigg[
2 (D-1) \, R_{ab}
- {\sum_{r = 2}^n}_{\, \,({\rm only} \, \, r \, \, {\rm even})}  \, r \, 
Q_{ab12 ... r-2} \, l^1 l^2 ... l^{r-2}
\bigg] \, l^a l^b = 0,
\; \; \; \; \; \; \; \; \forall l^a \, {\rm null}.
\end{eqnarray}

But
equation (\ref{isoq-15.2}) 
(in absence of $r$-odd terms
from the comment just above equation
(\ref{isoq-16.1}))
means

\begin{eqnarray}
(D-1) \, R_{ab} l^a l^b =
{\sum_{r = 2}^n}_{\, \,({\rm only} \, \, r \, \, {\rm even})}
\, Q_{12 ... r} \, l^1 l^2 ... l^r
+ \eta, 
\; \; \; \; \; \; \; \; \forall l^a \, {\rm null},
\end{eqnarray}
with $\eta = \eta(x)$ a scalar independent of $l^a$.
Crossing this with (\ref{isoq-16.4}),
gives 

\begin{eqnarray}
{\sum_{r = 2}^n}_{\, \,({\rm only} \, \, r \, \, {\rm even})}
\, Q_{12 ... r} \, l^1 l^2 ... l^r
+ \eta =
{\sum_{r = 2}^n}_{\, \,({\rm only} \, \, r \, \, {\rm even})}
\, \frac{r}{2}
\, Q_{12 ... r} \, l^1 l^2 ... l^r,
\; \; \; \; \; \; \; \; \forall l^a \, {\rm null}.
\end{eqnarray}

Apart from the trivial case in which all the terms are zero,
this equation can be satisfied only if there is one and only one
term not zero: that with $r = 2$ (also, implying $\eta = 0$)
(to be convinced, it suffices to look at what happens
if we send $l^a$ to $k \, l^a$, with $k$ a constant). 
This shows that we must have 
$
Q_* = Q_{ab} l^a l^b
$
and we are back to case $(c)$ of the main text,
i.e. case $(f)$ adds nothing to case $(c)$.


\end{document}